\documentclass[twocolumn, physrev, superscriptaddress,,amsmath,amssymb]{revtex4-2}

\usepackage{graphicx}
\usepackage{enumitem}
\usepackage{dcolumn}
\usepackage{ amssymb }
\usepackage{braket}
\usepackage{bm}
\usepackage{caption}
\usepackage{ragged2e}
\usepackage{hyperref}
\usepackage{comment}
\hypersetup{
    colorlinks=true,
    linkcolor=blue,
    filecolor=magenta,      
    urlcolor=blue,
    citecolor=blue,
}
\begin{document}

\preprint{APS/123-QED}

\title{\textbf{Efficient Photon Pair Generation from an Emitter in a Cavity} 
}%

\author{M.I. Mazhari}
\email{isamazhari20@iitk.ac.in}
\affiliation{Indian Institute of Technology Kanpur, Department of Physics}

\author{Rituraj}
\email{rituraj@iitk.ac.in}
\affiliation{Indian Institute of Technology Kanpur, Department of Electrical Engineering}

\date{\today}

\begin{abstract}
 Two-photon states are essential for fundamental applications in quantum information. One of the primary methods of two-photon generation is based on parametric down-conversion, but this suffers from low efficiency and a large footprint. This work presents a detailed theoretical investigation of an alternative approach: two-photon generation from an emitter in a doubly resonant cavity. The system is modeled by the Lindblad master Equation, and an approximate analytical solution is derived to determine the experimentally achievable limits on efficiency and brightness. Additionally, the optimal cavity parameters for achieving these limits are also identified. For experimentally feasible parameters, the maximum efficiency is approximately $35\%$, which is significantly higher than that of parametric down-conversion-based methods. The optimal rate and efficiency for two-photon generation are achieved when the outcoupling rate of the cavity mode at the two-photon emission frequency matches the single-photon atom-field coupling strength. Moreover, the outcoupling rate of the cavity mode at the one-photon emission frequency for single photons should be minimized. The cavity field properties are also examined by studying the second-order correlation function at zero time delay and the Mandel Q parameter, revealing highly bunched two-photon emission and super-Poissonian statistics. The quantum-jump framework, combined with Monte Carlo simulations, is used to characterize the mechanism of two-photon emission and the emission spectra of the cavity. Two-photon emission is demonstrated to be a rapid cascade process of quantum jumps, and its spectrum consists of three prominent peaks corresponding to transitions between the dressed states of the system.  

\end{abstract}
\maketitle

\section{Introduction}

Two-photon states are crucial for various fundamental applications in quantum information science. The commonly used method to generate them is based on parametric down-conversion\cite{Pan_2012, PhysRevLett.25.84,kwiat1999ultrabright,PhysRevLett.75.4337}, which suffers from low efficiency, with the maximum efficiency being around $5\%$\cite{couteau2018spontaneous}. Additionally, these systems are bulky, require a strong pump laser, and have poor mode selectivity, i.e., the photons are emitted into different modes.  This motivates us to study the two-photon generation from a promising alternative: an incoherently excited emitter in an appropriately designed cavity. This allows for the possibility of electronic pumping as well as better control over the modes in which photons are emitted.\\

Two-photon emission can occur when an electron in the excited state of an atom decays back to the ground state via a virtual transition to an intermediate state, releasing two quanta of energy. In free space, this process is dominated by single-photon emission\cite{tanoudji1998atom}, but a resonant cavity alters the spontaneous emission rates, leading to the Purcell enhancement of the otherwise weak two-photon spontaneous emission rate\cite{munoz_2014,munoz2015enhanced}. A recent study \cite{rituraj2023efficient} addresses the design of a simple doubly resonant photonic crystal cavity aimed at optimizing two-photon generation via Purcell enhancement of degenerate two-photon emission and the entrapment of higher-frequency photons. Consequently, an emitter coupled to a cavity/photonic structure with a suitably designed band structure and subjected to continuous incoherent pumping/excitation can act as an efficient two-photon source. We choose incoherent pumping because it is the standard mechanism used to ensure continuous photon generation in light sources and can be implemented easily using various methods. These include electrical injection\cite{del2010two,injection}, coupling to a gain medium/bath of inverted harmonic oscillators\cite{inverted_pump,inbath2}, coupling to a higher excited state from which the emitter rapidly decays to the upper level of the transition of interest\cite{incoh_mech_1,incoh_mech_2,incoh_mech_3}, etc.\\


\begin{figure*}[hbt!]
\includegraphics[width=140mm,scale=0.8]{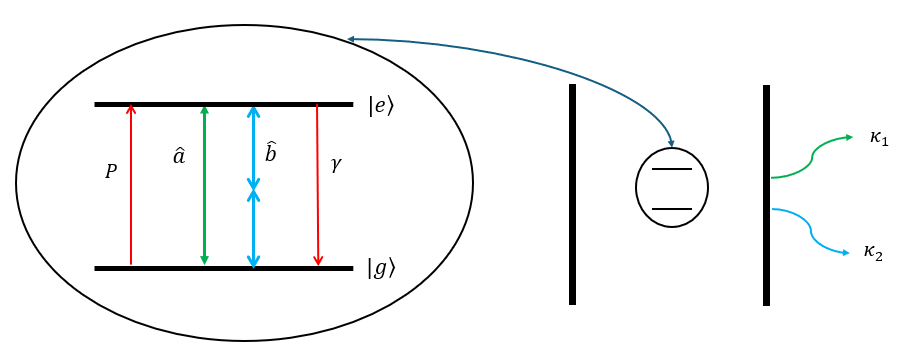}
\caption{\label{fig:2level} Diagram of the two-level system's energy levels and the transitions. The atom consists of ground and excited states denoted by $\ket{g}$ and $\ket{e}$ respectively, interacting with the $\omega_0$ (annihilation operator $\hat{a}$) and $\omega_0/2$ (annihilation operator $\hat{b}$) modes of the cavity which drive one and two-photon transitions respectively, denoted by bidirectional arrows. The unidirectional arrows represent incoherent excitation at rate $P$ and decay of the excited state at rate $\gamma$. The outcoupling rates of the $\omega_0,\omega_0/2$ modes are $\kappa_1,\kappa_2$ respectively.}
\end{figure*}

In this work, we examine the two-photon generation from a system consisting of an emitter coupled to such a doubly resonant cavity/resonator. The two resonant modes of the cavity separately drive one and two-photon transitions between the energy levels. We derive an approximate analytical solution of the Lindblad master equation by making a manifold approximation which is verified using numerical simulations. We use the analytical solution to determine the maximum efficiency and rate of two-photon emission, and also identify the optimal values for the cavity parameters. This combined analytical–numerical framework allows us to identify a previously unreported efficiency maximum which occurs when the two-photon outcoupling rate equals the one-photon vacuum rabi frequency. We further analyze the cavity-field statistics, the emission spectrum and underlying quantum-jump dynamics using Monte Carlo simulations\cite{crowder2020quantum, quant_traject, inverted_pump}. The Monte Carlo simulations reveal the mechanism of two-photon emission to be a fast cascade process of quantum jumps.\\

 This paper is organised as follows: Section \ref{sec:section2} describes the theoretical formulation of the system model. In section \ref{sec:section3}, the two-photon generation efficiency and emission rates are defined, and the approximate analytical solution is derived using the  manifold approximation. Efficiency, emission rates, and the cavity field statistics and their dependence on the system's parameters, are studied in section \ref{sec:sectionSSR}. The optimal cavity parameters are identified along with the practically achievable limits of efficiency. In Section \ref{sec:section6}, the mechanism of two-photon emission and the emission spectra are analysed using the quantum-jump framework and Monte-Carlo simulations. The appendix \ref{sec:derivation} describes one possible realization of our system model as an effective Hamiltonian of a three-level system and delineates the region in the parameter space where the effective two-level system model is a valid approximation. Appendix \ref{sec:SSE} details the calculations and the approximations required to arrive at closed-form expressions of the system's steady-state statistics.

\section{\label{sec:section2} System Model}

The system consists of an emitter with ground state $\ket{g}$ and an excited state $\ket{e}$, coupled to two electromagnetic modes of a doubly resonant cavity, as shown in Figure \ref{fig:2level}. The energy difference between the excited state and the ground state is $\omega_0$ ($\hbar=1$). The two cavity modes, at frequencies $\omega_0/2$ and $\omega_0$, resonantly couple to the two-photon and one-photon transitions, respectively. The Hamiltonian for the system is as follows:

\begin{equation} \label{Ham}
    H=H_0 + H_I, 
\end{equation}

where $H_0$ is the Hamiltonian for the uncoupled emitter-cavity system and $H_I$ is the interaction Hamiltonian, given by 
\begin{equation}
     H_0 = \frac{1}{2} \omega_0 \sigma_z  + \omega_0 a^{\dagger}a +\frac{\omega_0}{2}b^{\dagger}b,
\end{equation}

and 
\begin{equation}
 \label{HI}   H_I=\underbrace{g_1(a^{\dagger}\sigma_- + a\sigma_+)}_{H_{I1}} +  \underbrace{g_2(b^{\dagger 2}\sigma_- + b^2\sigma_+)}_{H_{I2}}.
\end{equation}

Here, $\sigma_z=\ket{e}\bra{e}-\ket{g}\bra{g}$, and $\sigma_-=\ket{g}\bra{e}$, $\sigma_+=\ket{e}\bra{g}$ are the emitter lowering and raising operators respectively. $\hat{a}$ and $\hat{b}$ are the annihilation operators of the cavity modes at $\omega_0$ and $\omega_0/2$ respectively. $g_1$ and $g_2$ are the coupling strengths between the emitter and the $\omega_0$ and $\omega_0/2$ modes respectively. 
$H_I$ describes a generalized Jaynes-Cummings interaction \cite{singhgeneral} between the emitter and the field consisting of both one-photon $(H_{I1})$ and two-photon $(H_{I2})$ transitions between the ground and excited states.

Note that the $H_{I2}$ term in the interaction Hamiltonian is a phenomenological term that is used to model the simultaneous absorption or emission of two photons by the emitter \cite{ZubairyTPEValid,agarwaltwoph,alsing_two_ph,puri1988_two_ph_eff,two_ph_Ham_finite_Q,micromaser}. We propose one possible physical realization of this phenomenological model by using a three-level emitter interacting with the two resonant cavity modes in Appendix \ref{sec:derivation}. In this case, the Hamiltonian of Eq. \ref{Ham} serves as an effective Hamiltonian derived by performing a suitable unitary transformation on the Hamiltonian of the three-level system which eliminates the additional third level in the limit of large detuning. The derivation of this Hamiltonian and the delineation of the parameter space where our phenomenological model is a valid approximation of the three-level system are presented in Appendix \ref{sec:derivation}. 

We work in the Fock state basis where $\ket{i,j,k}\equiv \ket{i}\otimes\ket{j}\otimes \ket{k}$ represents the system state with the emitter in state $i\in{\{g,e\}}$, $j$ photons in the mode at $\omega_0/2$ and $k$ photons in the $\omega_0$ mode. The state of the system is described by a density matrix with the diagonal and off-diagonal terms representing the populations and the coherence in the above basis.\\

We consider the open system dynamics described by the Lindblad master equation\cite{breuer2002theory,carmichael2009open,del2010two}, in terms of the system density matrix $\rho$:
\begin{equation} \label{Lindblad2}
    \frac{d\rho}{dt}=-i[H,\rho]+\frac{\kappa_1}{2}\mathcal{L}_{a}\rho + \frac{\kappa_2}{2}\mathcal{L}_{b}\rho +\frac{P}{2}\mathcal{L}_{\sigma_+}\rho+\frac{\gamma}{2}\mathcal{L}_{\sigma_-}\rho
\end{equation}

where $\mathcal{L}_c$ is the Lindblad super-operator for the system jump (or collapse) operator $c$ described by
\begin{equation}
\label{Lind}
    \mathcal{L}_c\rho=2c \rho c^{\dagger}-c^{\dagger}c\rho-\rho c^{\dagger}c.
\end{equation}
In contrast to the term involving the Hamiltonian $H$ in Eq.\ \ref{Lind}, the jump operators result in non-unitary evolution of the system because of its interaction with the environment. The four jump operators in the master equation \ref{Lind} correspond to the following processes.
\begin{enumerate}
\item The photons in the cavity modes at $\omega_0$ leak out at a rate $\kappa_1$ described by the jump operator $\sqrt{\kappa_1}a$.
\item The photons in the cavity modes at $\omega_0/2$ leak out at a rate $\kappa_2$ described by the jump operator $\sqrt{\kappa_1}b$.
  \item The Lindblad term with the jump operator $\sqrt{P}\sigma_+$ represents the incoherent excitation of the emitter at a constant rate $P$ to ensure a continuous generation of photons in steady state. 
   
  \item The decay of the excited state due to radiative emission into the non-cavity modes and other non-radiative processes at rate $\gamma$ is described by the jump operator $\sqrt{\gamma}\sigma_-$.
\end{enumerate}

\section{\label{sec:section3}Theoretical Analysis}

The steady state of the system is defined by:
\begin{equation}
    \frac{d\rho}{dt}=i[H,\rho]+\frac{\kappa_1}{2}\mathcal{L}_{a}\rho + \frac{\kappa_2}{2}\mathcal{L}_{b}\rho +\frac{P}{2}\mathcal{L}_{\sigma_+}\rho+\frac{\gamma}{2}\mathcal{L}_{\sigma_-}\rho=0
\end{equation}
The steady state expectation values of different operators can then be determined from the steady state density matrix using the formula $\frac{d \langle c \rangle}{dt}=Tr[c\frac{d\rho}{dt}]=0$. This allows us to derive the rate equations and relations between various other expectation values, one of which is:

\begin{eqnarray} \label{Eq_eq}
    P \langle \ket{g}\bra{g} \rangle &=& \kappa_1 \langle a^{\dagger}a \rangle + \frac{\kappa_2}{2} \langle b^{\dagger}b \rangle + \gamma \langle \ket{e}\bra{e} \rangle
\end{eqnarray}

This rate equation describes a steady-state balance between the rate at which the system gains and loses "excitations", due to the Lindblad jump operators. In order to understand this further, we define an excitation number operator $\hat{N}$ as the sum of excited state population, the number of $\omega_0 $ photons and the number of $\omega_0/2$ photon pairs: 
\begin{equation}
    N=\ket{e}\bra{e} +a^{\dagger}a+\frac{b^{\dagger}b}{2}.
\end{equation}

Note that there is a factor of $0.5$ for $\omega_0/2$ mode because its interaction with the atom  involves simultaneous exchange of two $\omega_0/2$ photons. The operator $\hat{N}$ has eigenstates given by $\ket{g,j,k}$ and $\ket{e,j,k}$, with eigenvalues $(j+k/2)$ and $(j+1+k/2)$ respectively, i.e.,

\begin{align}
    \nonumber\hat{N}\ket{g,j,k}=(j+k/2)\ket{g,j,k}\\ \hat{N}\ket{e,j,k}=(j+1+k/2)\ket{g,j,k}
\end{align}

The excitation operator commutes with the Hamiltonian, i.e., $[H,\hat{N}]=0$, implying that the number of excitations is a conserved quantity for the evolution of the closed system. However, for the open system, $d\hat{N} /dt \neq 0$, as couplings to the environment cause the system to gain or lose excitations. In steady state, the rate of excitation balances the rate of de-excitation, leading to $d \langle N \rangle/dt=0$.
The left-hand side of Eq \ref{Eq_eq} represents the excitation of the system, where the incoherent pump excites the emitter from the ground to the excited state. The terms on the right-hand side represent de-excitation due to the cavity outcoupling and emitter decay. $\kappa_1 \langle a^{\dagger}a \rangle$ and $\frac{\kappa_2}{2} \langle b^{\dagger}b \rangle$ represent the rates of one-photon emission (OPE) and two-photon emission (TPE) from the cavity, respectively. The term $\gamma \langle \ket{e}\bra{e} \rangle$ is the decay rate of emitter excitation due to emission into the free space modes and non-radiative processes. \\

This is also corroborated by the quantum jump formalism \cite{MCWF,PlenioQJump,molmer1993monte,Jump_Gardiner}, which describes the system's evolution as coherent periods of evolution interspersed with random quantum jumps where the system abruptly transitions to a different state. The coherent evolution is governed by the Schrodinger equation with an effective non-Hermitian Hamiltonian $H_{e}$ (Eq.\ \ref{H_eff}), and the random quantum jumps are determined by the jump operators corresponding to the ones in the Lindblad master equation. The average of many such trajectories, consisting of coherent evolutions and random jumps, reproduces the results obtained from the Lindblad master equation.

\begin{equation} \label{H_eff}
   H_{e}=H-\frac{i}{2}(\kappa_1 a^{\dagger}a+\kappa_2b^{\dagger}b+P\ket{g}\bra{g}+\gamma \ket{e}\bra{e}) 
\end{equation}

 As $[H_{e},N]=0$, the number of excitations is conserved during the coherent evolution and changes only when quantum jumps occur. The collapse operators $\sqrt{\kappa_1}a,\sqrt{\kappa_2}b,\sqrt{\gamma}\sigma_-$ cause quantum jumps to states with lower $\langle N \rangle$. At steady state, these are balanced out by quantum jumps to states with higher $\langle N \rangle$, caused by $\sqrt{P}\sigma_+$. This balance is represented by Eq. \ref{Eq_eq}.\\

Hence, we define the efficiency of two-photon emission $\eta$ as the ratio of the TPE rate and the rate of pumping the emitter to the excited state:

\begin{equation} \label{eff_expr}
    \eta=\frac{\kappa_2 \langle b^{\dagger}b \rangle }{2 P \langle \ket{g}\bra{g} \rangle } \times 100
\end{equation}

We denote the TPE rate, OPE rate and decay rate as $T,O,L$ respectively:

\begin{align}
    T&= \frac{\kappa_2\langle b^{\dagger}b \rangle}{2} \label{TPE} \\
    O&= \kappa_1 \langle a^{\dagger}a \rangle  \label{OPE}\\
    L&= \gamma \langle \ket{e}\bra{e} \rangle \label{loss} 
\end{align}

The relationships between the expectation values of various operators form an infinite series that cannot be solved analytically \cite{scully1997quantum}. Thus, in order to obtain a solution, we must make a series of approximations as discussed below.

\subsection{\label{sec:ManifoldApprox}Manifold Approximation}

We define a manifold $M_n$ as the subspace spanned by the eigenstates of $\hat{N}$ with eigenvalue $n$. The basis states of the first three manifolds are listed as:
\begin{description} [itemsep=1pt]
    \item [\textbf{$M_{0}$}] $\ket{g,0,0}$ 
    \item [\textbf{$M_{0.5}$}] $\ket{g,0,1}$ 
    \item [\textbf{$M_{1}$}] $\ket{e,0,0},\ket{g,1,0},\ket{g,0,2}$ 
\end{description}


As discussed above, the system's time evolution consists of coherent evolution within a particular manifold $M_n$, interspersed with quantum jumps to manifolds $M_k$ where $k\neq n$. 

At high pumping rate, the system cycles through manifolds with large values of $n$. This makes the system intractable to solve as the higher manifolds have higher dimensions. However, as we show later, a higher $P$ leads to a lower TPE efficiency. This motivates us to investigate the low pump regime.

At low $P$, the system predominantly remains in manifolds with small values of $n$.  
In particular, we consider sufficiently small values of $P$ such that the system seldom excites to states lying beyond $M_1$. We include manifolds up to $M_1$ because that is the lowest excitation manifold where the system can undergo coherent two-photon transitions. These transitions are indicated by the oscillations of the probability amplitudes of the states $\ket{e,0,0},\ket{g,0,2}$ during the coherent evolution periods (with no quantum jumps) due to $H_{e}$. We follow a two-step dimension reduction procedure. First, we restrict the Hilbert space of the system to the basis states where the number of photons in the $\omega_0$ and $\omega_0/2$ modes are at most $1$ and $2$ respectively. This results in 18 equations in terms of 18 steady-state operator expectation values, which are listed in Appendix \ref{sec:SSE}. We then restrict the Hilbert space to the manifolds $M_0, M_{0.5}, M_1$.  In the restricted Hilbert space, for any operator \(\hat{c}\),
\begin{equation}
    \langle c \rangle=Tr[c\rho]=\sum_{i=1}^5\bra{i} c \rho \ket{i}
\end{equation}


Here, the trace involves sum over the five basis states of the manifolds $M_0,M_{0.5}$ and $M_1$. This results in several additional relations between the steady-state expectation values, which are listed in the Appendix \ref{sec:SSE}. This allows us to obtain closed-form expressions for the efficiency and photon emission rates.

\begin{widetext} 
    \begin{align} 
        T = \frac{\kappa_2}{2} \langle b^{\dagger}b \rangle &= \frac{2 \kappa_2 g_2 \nu P}{\frac{\phi}{\xi}(2\kappa_1g_1 +(\gamma +P)\xi)+2g_2\nu^2\kappa_2} \hspace{1cm} \label{MT} \\
        O = \kappa_1 \langle a^{\dagger}a \rangle &= \frac{2g_1\kappa_1 P(\phi-\kappa_2g_2\nu/(\kappa_2+\kappa_1/2))}{\phi(\xi(P+\gamma)+2\kappa_1g_1)+2g_2\nu^2\kappa_2\xi} \label{MO} \hspace{1cm}
    \end{align}
    \begin{align}
        L = \gamma \langle \ket{e} \bra{e} \rangle &= \frac{\gamma P\phi}{\frac{\phi}{\xi}(2\kappa_1g_1 +(\gamma +P)\xi)+2g_2\nu^2\kappa_2} \label{ML}\hspace{1cm}  \\
        \eta &= \frac{2\kappa_2g_2\nu}{\frac{\phi}{\xi}(2\kappa_1g_1 +\gamma\xi)+2g_2\nu^2\kappa_2\frac{2\kappa_1g_1 +\gamma\xi}{2\kappa_1g_1 +(\gamma +P)\xi}} \times 100 
        \label{eff}
    \end{align}
\end{widetext}
\newpage
Here, 
\begin{align}
    \xi &= 2g_1+\frac{2\kappa_1g_2^2}{(\kappa_2+\kappa_1/2)g_1}+\frac{\kappa_1(\kappa_1+P+\gamma)}{2g_1} \\
    \nu&=1-\frac{\kappa_1g_1}{(\kappa_2+\kappa_1/2)\xi} 
\end{align}
\begin{align}
    \phi &= \frac{\kappa_2}{2g_2}(\kappa_2+P/2+\gamma/2+\frac{2g_1^2}{2\kappa_2+\kappa_1}) - \frac{\kappa_2\kappa_1g_1g_2}{(\kappa_2+\kappa_1/2)^2\xi} 
\end{align}



\begin{figure*}[hbt!]
\includegraphics[width=150mm,scale=0.8]{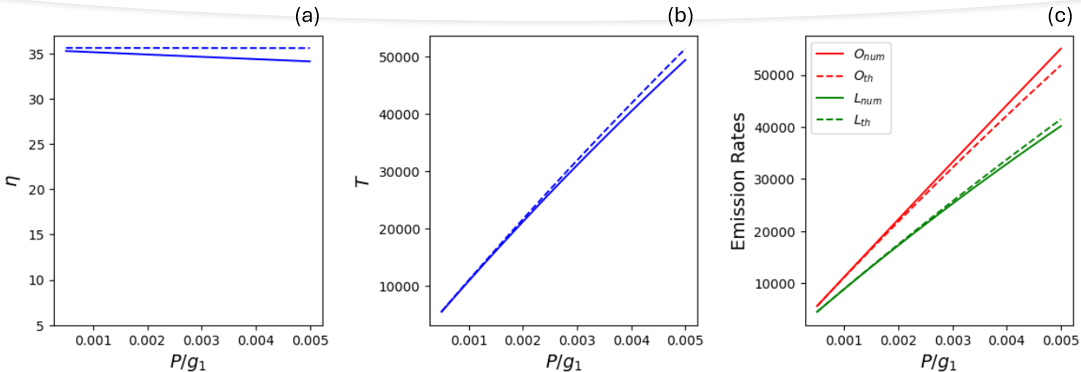}
\caption{\label{fig: Pg1} Dependence of the steady state statistics on $P$, for $\kappa_1=0.02g_1,\kappa_2=g_1,\gamma=0.016g_1$. (a) depicts $\eta$, (b) depicts $T$, and (c) depicts $O$ and $L$ with solid and dashed curves showing the simulation and analytical results respectively.}
\end{figure*}

\begin{figure*}[hbt!]
\includegraphics[width=150mm,scale=0.8]{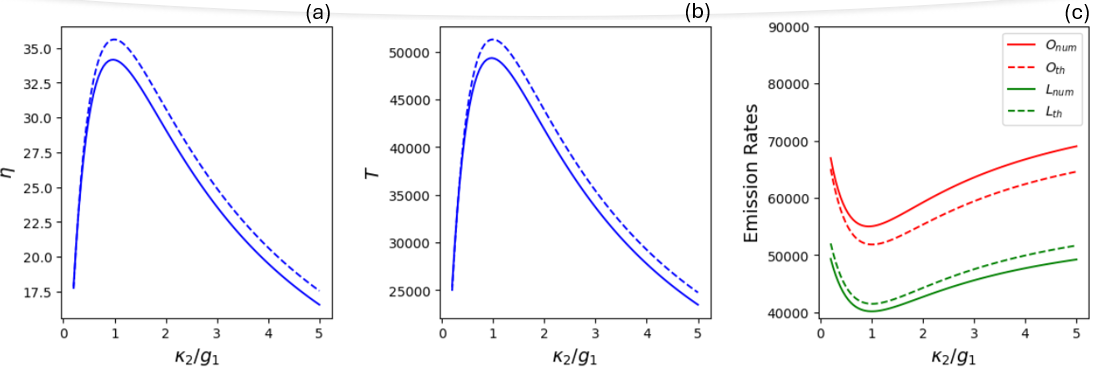}
\caption{\label{fig: kappa2g1} Dependence of the steady state statistics on $\kappa_2$, for $\kappa_1=0.02g_1,P=0.005g_1,\gamma=0.016g_1$. (a) depicts $\eta$, (b) depicts $T$, and (c) depicts $O$ and $L$ with solid and dashed curves showing the simulation and analytical results respectively.}
\end{figure*}

\section{\label{sec:sectionSSR}Steady State Results}

Having obtained the steady-state solution analytically, we now study the magnitudes of $\eta$, the emission rates, the second-order correlation function at zero time delay ($g^2(0)$), and the Fano Factor. We also compare the analytical model with exact numerical simulations performed in QuTiP\cite{qtip}. This numerical method directly solves the Master equation at steady-state using sparse LU decomposition of the system's Liouvillian. We choose the values of the parameters considering a superconducting circuit QED setup\cite{cyclic3,cyclic3_3,cyclic3_2} to realize our model, as discussed in Appendix \ref{sec:derivation}. The values of the $g_1$, $\omega_0$ and $\gamma$ are taken from Ref. \cite{cyclic_params}. We select a Q-factor value i.e., $Q=\omega_0/\kappa$, of approximately \(10^{5}\) for the \(\omega_0\) mode, ,which is typical for a metamaterial resonator\cite{Qfactor,Qfactor2}. This can be engineered to be doubly resonant, similar to the photonic crystal cavity\cite{rituraj2023efficient}, and plays the role of the cavity. We consider the resonator $Q$ values in the range [$10^{-6}\omega_0$,$10^{-4}\omega_0$] i.e., a $\kappa_1$ range of [$\kappa_1=0.02g_1$,$\kappa_1=2g_1$]. The value of $g_2$ is taken as $g_2=0.1g_1$, for which our phenomenological model is valid and gives highest efficiency, as discussed in Appendix \ref{sec:derivation}.  \\

The analytical and numerical values of \(\eta\), T, O and L are plotted with respect to the pump rate $P$ and the cavity leakage rates $\kappa_2$ and $\kappa_1$ in figures 2, 3, and 4 respectively. The solid lines represent the analytical values calculated using Eqs. \ref{MT},\ref{MO},\ref{ML},\ref{eff}, while the dashed lines indicate the numerical values obtained from QuTip simulations. We discuss these results next. 

\subsection{\label{sec: Opt}Optimal Cavity Parameters}

For the chosen range of cavity parameters, i.e. $\kappa_1,\gamma <<g_1$ and at low $P$, the analytical expressions can be simplified to:  
\begin{align} 
        \eta &\approx \frac{4\kappa_2g_2^2}{(\gamma+\kappa_1)(\kappa_2^2+g_1^2)} \times 100  \hspace{1cm} \label{eta_P}  \\
        \frac{\kappa_2}{2} \langle b^{\dagger}b \rangle &\approx \frac{4g_2^2P\kappa_2}{(\gamma+\kappa_1)(\kappa_2^2+g_1^2)} \label{T_P} \\
        \kappa_1 \langle a^{\dagger}a \rangle &\approx \frac{\kappa_1 P}{(\gamma+\kappa_1)} \label{O_P}  \\
        \gamma \langle \ket{e} \bra{e} \rangle &\approx \frac{\gamma P}{(\gamma+\kappa_1)} \label{L_P}
\end{align}

Figure \ref{fig: Pg1} shows the efficiency and emission rates as a function of $P$. Figure 2a shows that the efficiency is almost independent of $P$. Consequently, the TPE rate increases linearly with $P$ as shown in Figure 2b. This allows for an enhancement of the two-photon emission rate by increasing $P$. However, at still higher $P$, the efficiency starts to decline, and the deviation between the analytical and numerical values increases, due to the increased contribution of higher order manifolds in the system dynamics. Thus, high efficiency occurs in the low $P$ regime where the analytical model is a good approximation as can be seen from the close agreement with the simulation results. The OPE and atomic loss rates $O,L$ are shown in Figure 2c by blue and green curves respectively and are proportional to $P$. Note that, the atomic loss rate has a fundamental upper bound of $\gamma$ and thus at high $P$, the OPE would be the dominant process contributing to a loss in TPE efficiency.   \\

The dependence of efficiency and emission rates on the decay rate of $\omega_0/2$ mode (\(\kappa_2\)), is illustrated in Figure \ref{fig: kappa2g1}. As shown in figures 3(a) and 3(b) respectively, the TPE rate and efficiency $\eta$ first increase with $\kappa_2$, achieve maximum values around $\kappa_2 \approx g_1$, and then decrease with further increase in $\kappa_2$.
This can be explained as follows. At very low values of $\kappa_2$, the cavity mode in $\omega_0/2$ is almost closed, leading to a low TPE rate and efficiency. For very large values of $\kappa_2$, the effective density of states around $\omega_0/2$ becomes small, again resulting in a poor TPE efficiency. This is also in agreement with the analytical model Eqs.(\ref{eta_P},\ref{T_P}), which predicts that the $\eta$ and the TPE rate are maximized when $\kappa_2=g_1$. 
The peak value of efficiency is observed to be $35\%$, which is several times higher than the efficiency of SPDC \cite{SPDC_eff}. This peak is asymmetric around the optimal point, as $\eta$ decreases slowly for $\kappa_2>g_1$, and sharply for $\kappa_2<g_1$. The OPE and the atomic loss rates as a function of $\kappa_2$ are plotted in figure 3(c) by blue and green curves, respectively. Both the curves have a minima around $\kappa_2=g_1$. Furthermore, the minima of the atomic loss rate also corresponds to the maxima of the population in the ground state ($\langle \ket{g}\bra{g} \rangle$). Thus, an optimal two-photon generation rate is achieved at $\kappa_2 \approx g_1$, and coincides with the simultaneous enhancement of the two-photon emission efficiency rate and the ground state population, and suppression of the one-photon emission and emitter decay rate.\\

\begin{figure*}[hbt!]
\includegraphics[width=150mm,scale=0.8]{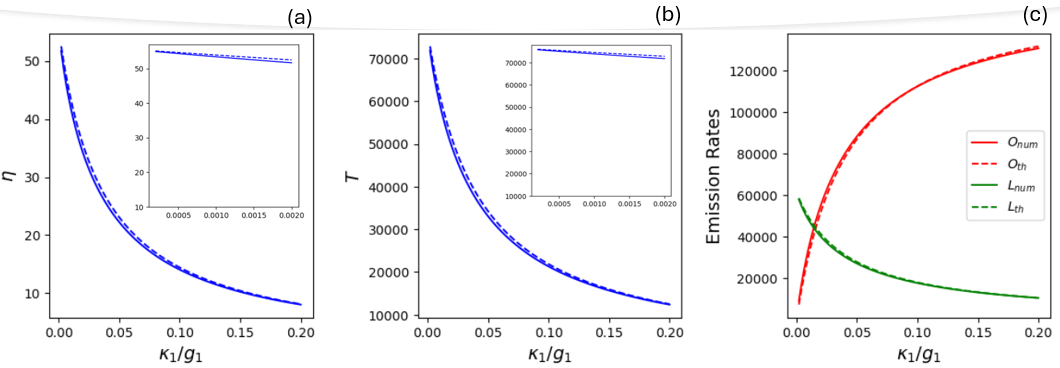}
\caption{\label{fig: kappa1g1}Dependence of the steady state statistics on $\kappa_1$, for $\kappa_2=g_1,P=0.005g_1,\gamma=0.016g_1$. (a) depicts $\eta$, (b) depicts $T$, and (c) depicts $O$ and $L$ with solid and dashed curves showing the simulation and analytical results respectively.}
\end{figure*}

The dependence of efficiency and various emission rates on the loss rate of cavity mode at $\omega_0$ ($\kappa_1$) is shown in Figure \ref{fig: kappa1g1}. The range of $\kappa_1$ has taken to be from $\kappa_1=0.002g_1$ to $\kappa_1=0.2g_1$, corresponding to the quality factor ranging from $10^6$ to $10^4$, respectively. Efficiency (Fig. \ref{fig: kappa1g1}(a)) and TPE rate (Fig. \ref{fig: kappa1g1}(b)) increase as $\kappa_1$ decreases due to a reduction in the losses arising from photon emission at $\omega_0$ and the photon recycling process, i.e., reabsorption of photons at $\omega_0$ followed possibly by emission as bi-photons \cite{rituraj2023efficient}. The efficiency increases from $10\%$ at $\kappa_1=0.2g_1$ to more than $50\%$ at $\kappa_1=0.002g_1$, and the TPE rate also increases in a similar manner. From Eqs. \ref{eta_P},\ref{T_P}:

\begin{align}
    \eta,T \propto \frac{1}{\kappa_1+\gamma}
\end{align}

Thus, decreasing $\kappa_1$ improves the TPE efficiency significantly when $\kappa_1>\gamma$ and for $\kappa_1<<\gamma$, the TPE efficiency is limited by the atomic loss rates. This is also evident in the dependence of OPE and atomic loss rates on $\kappa_1$, as shown by the blue and green curves respectively in Figure \ref{fig: kappa1g1}(c). The OPE rate decreases with a decrease in $\kappa_1$ despite an increase in the mean photon number $\langle a^\dagger a\rangle$ in the cavity mode at $\omega_0$. An increase in the mean cavity photon number also results in an increase in the excited state population and thus an increase in the atomic loss rate. For sufficiently small values of $\kappa_1$, the atomic loss rates become larger than the OPE rate.

Therefore, optimal two-photon generation simultaneously requires low values for both $\kappa_1$ and $\gamma$. This is shown in the insets of Figure \ref{fig: kappa1g1}(a) and (b), where $\eta$ and $T$ are plotted at low values of $\kappa_1$ such that $\kappa_1<<\gamma$. In this case, even though $\kappa_1$ decreases by two orders of magnitude, the increase in $\eta$ and the TPE rate is insignificant. The maximum possible efficiency $\eta_{max}$ that can be obtained for a given $\gamma$ and for optimal $\kappa_2$, i.e., $\kappa_2=g_1$ is
\begin{align}
    \eta_{max}= \lim_{\kappa_1 \rightarrow 0}\eta \big|_{\kappa_2=g_1}=\frac{2g_2^2}{\gamma g_1} \approx 55.21 \%
\end{align}

Thus, we have now determined the optimal cavity parameters for realizing an efficient two-photon emission. Namely, the two-photon out-coupling rate $\kappa_2$ should be equal to $g_1$ and the one-photon out-coupling rate $\kappa_1$ should be smaller than the atomic decay rate $\gamma$. For these optimal cavity parameters, the TPE efficiency is maximized at low pumping rates and the TPE rate increases linearly with $P$ without significant change in efficiency. Additionally, in this low $P$ regime, there is excellent agreement between the analytical model and numerically simulated results.

\subsection{\label{sec: FieldStat}Field Statistics}

Assuming an optimally designed cavity as discussed above, we now examine the Mandel Q Parameter and the normalized second-order correlation function at zero time delay $(g^2(0))$ for the $\omega_0 / 2$ mode, in order to analyze the statistical properties of the cavity and the output field.\\

The Mandel Q Parameter is defined as the ratio of the variance in the photon count and the mean i.e.,
\begin{equation}
    Q=\frac{\langle b^{\dagger 2}b^2 \rangle-\langle b^{\dagger}b \rangle^2}{\langle b^{\dagger}b \rangle}
\end{equation}
The value of $Q$ quantifies the deviations of the cavity field statistics from the Poissonian statistics. A positive value of Q indicates super-Poissonian statistics whereas a negative $Q$ implies sub-Poissonian statistics. $g^2(0)$ is defined as a measure of two-photon coincidence detection probability.
\begin{equation}
    g^2(0)=\frac{\langle b^{\dagger 2}b^2 \rangle}{\langle b^{\dagger}b \rangle^2}
\end{equation}

$g^2(0)>1$ corresponds to "bunched" light which, for our system, indicates that the two-photons are emitted in pairs in a short period of time.

From the steady state equations (\ref{sec:SSE}), $\langle b^{\dagger 2}b^2 \rangle = \frac{1}{2} \langle b^{\dagger}b \rangle$ in the low manifold approximation. Hence,
\begin{align} 
    Q =&\frac{0.5 \langle b^{\dagger}b \rangle -\langle b^{\dagger}b \rangle^2}{\langle b^{\dagger}b \rangle} \approx 0.5 \\
    g^2(0) &\approx \frac{1}{2 \langle b^{\dagger}b \rangle} \propto \frac{1}{P} \label{g20_eq}
\end{align}

\begin{figure*}[hbt!]
\includegraphics[width=150mm,scale=0.8]{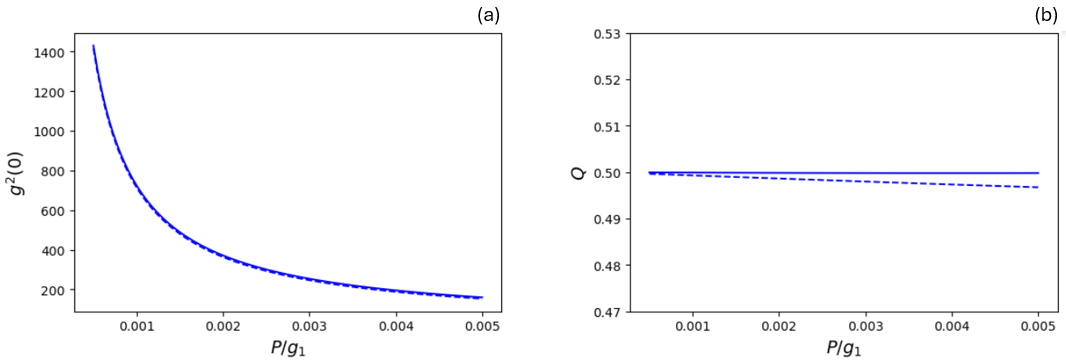}
\caption{\label{fig:g2F} Dependence of $g^2(0)$ and $Q$ on $P$, at $\kappa_2=g_1,\kappa_1=0.02g_1,\gamma=0.016g_1$}. 
\end{figure*}

\begin{figure*}[hbt!] 
\includegraphics[width=150mm,scale=0.8]{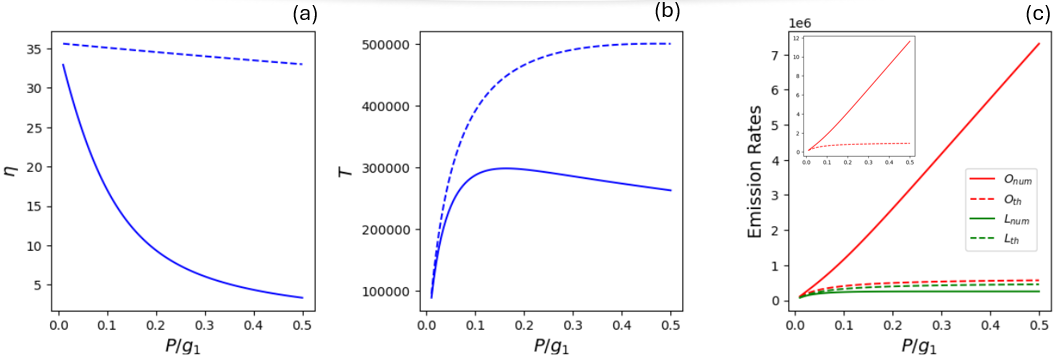}
\caption{\label{fig: highP} Efficiency and Emission Rates at high $P$, $\kappa_1=0.02g_1,\kappa_2=g_1,\gamma=0.016g_1$.Sub-figure (a) depicts $\eta$, Sub-figure (b) depicts $T$ and $L$. Sub-figure (c) depicts $O$. The inset of Sub-figure (c) depicts the number of photons i.e., $\langle a^{\dagger}a \rangle$}
\end{figure*}

Figure \ref{fig:g2F}(a) and \ref{fig:g2F}(b) plot the numerically simulated values (solid line) and analytical values (dashed line) of \( g^2(0) \) and $Q$ respectively as functions of the pump power \( P \). $Q$ remains approximately constant at 0.5, consistent with the theoretically predicted value, indicating a super-Poissonian statistics. The value of \( g^2(0) \) is of the order of a few hundred, indicating bunching and a high likelihood of coincidence detection. The magnitude of $g^2(0)$ decreases sharply as the pump rate increases, due to it being inversely proportional to $P$ (Eq. \ref{g20_eq}). Therefore, in order to obtain highly bunched photon pairs, it is desirable to have a low pump rate, but this comes at the cost of reduced biphoton emission rate, as discussed earlier in Section \ref{sec:sectionSSR}.

\subsection{\label{sec: HighPump}High Pump Regime}

For the optimal cavity parameters, we now investigate the high $P$ regime. The values of $\eta$ and the emission rates are plotted as a function of $P$ in Figures \ref{fig: highP}(a) and \ref{fig: highP}(b) respectively. The $P$ is varied from from $P=0.001g_1$, where $P<<\kappa_1,\kappa_2,\gamma$,  to \(P = 0.5 g_1\). The efficiency $\eta$ decreases significantly as P increases and its value deviates significantly from that predicted by the analytical expressions at high $P$, as shown in Figure \ref{fig: highP}a. Additionally, as shown by the blue curve in Figure \ref{fig: highP}b, the TPE rate peaks around $P \approx 0.15g_1$, reaching approximately $3 \times 10^{5}$ photon pairs emitted per second, with an efficiency of about $12\%$, which is more than twice that of SPDC \cite{SPDC_eff}. As the pump rate increases further, the TPE rate starts to decrease due to a significant decrease in efficiency. The TPE rates and the atomic loss rates become similar in magnitude at high $P$. Unlike the large inaccuracy observed in the efficiency values obtained from the analytical model at high $P$, the analytical model predicts the TPE rate and the atomic loss rate within a factor of $2$ of the correct values obtained from the simulation results even at high $P$. Figure \ref{fig: highP}(c) plots the OPE rate and shows that at high $P$, most of the power is dissipated as photon emission at $\omega_0$. The magnitude of OPE is about an order of magnitude higher than the TPE rate at high $P$. The large deviation between the OPE rates obtained from the analytical model and the numerical simulation points to the primary reason for the failure of the analytical model in predicting $\eta$ at high $P$. \\

As discussed before, the low manifold approximation fails at high $P$ as the system cycles through higher-order manifolds. This is evident from the the inset of Figure \ref{fig: highP}(c), which depicts the mean number of cavity photons $\langle a^{\dagger}a \rangle$ being greater than 1 for $P$ of the order of $10^{-1}g_1$. As discussed in \ref{sec:ManifoldApprox}, the system's probability of undergoing two-photon transitions between $\ket{g}$ and $\ket{e}$ decreases in higher manifolds, and one-photon transitions become predominant. This results in an increased number of intracavity $\omega_0$ photons and leads to most of the power being dissipated as one-photon emission, resulting in low $\eta$. It also leads to the discrepancy between the analytical and numerical values, which is most significant for $O$ and consequently, $\eta$. \\

Thus, high efficiency occurs in the low $P$ regime where $\langle a^{\dagger}a \rangle $ is small. This is also the regime where our analytical model is valid, and which has been analyzed in detail in the previous subsections \ref{sec: Opt} and \ref{sec: FieldStat}. Thus, for the remainder of the paper, we only consider the low $P$ regime.

\begin{figure*}[hbt!]
\includegraphics[width=150mm,scale=0.8]{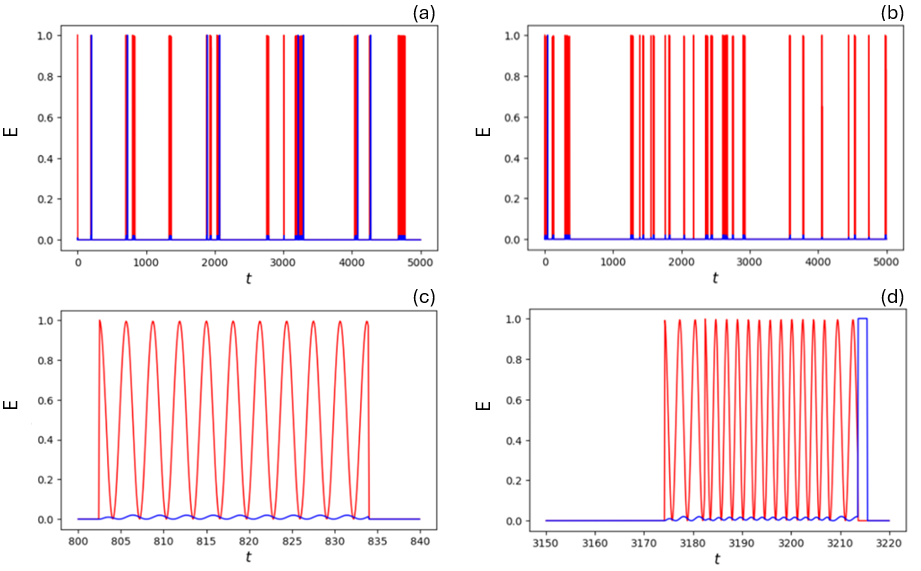}
\caption{\label{MC2} Subfigures (a) and (b) depict a single Monte Carlo trajectory of the system for the low and high efficiency cases respectively. The value of $\kappa_2=g_1,P=0.005g_1,\gamma=0.016g_1$ for both subfigures. For the subfigure on the left, $\kappa_1=0.2g_1$, whereas for the subfigure on the right, $\kappa_2=0.02g_1$.  Subfigure (c) shows a close-up view of a quantum jump resulting in one-photon emission or the decay of the excited state due to other losses, while subfigure (d) depicts a close-up view of the two-photon emission.} 
\end{figure*}

\section{\label{sec:section6}Quantum Jump Analysis}

We now use the Quantum Jump formalism and Monte Carlo simulations, as introduced in subsection \ref{sec:section3}, to better understand the mechanism of the two-photon emission and to study the two-photon emission spectrum.

\subsection{Mechanism of Emission}

As discussed in section \ref{sec:section3}, in the Monte-Carlo wavefunction formalism \cite{quant_traject,inverted_pump}, a single quantum trajectory consists of coherent evolution of the state $\ket{\psi(t)}$ under the influence of $H_{e}$ (Eq. \ref{H_eff}), which is interspersed with random quantum jumps. We discretize the time evolution of the system in multiple time steps of size $\delta t$. At time $t$, the system undergoes one of the four possible quantum jumps with corresponding probabilities $\delta p_j=\braket{\psi(t)|c_j^{\dagger}c_j|\psi(t)}\delta t$. With a much higher probability $(1-\Sigma_j\delta p_j)$ it evolves under $H_{e}$ as $\ket{\psi(t+\delta t)}=e^{-iH_e\delta t}\ket{\psi(t)}/\sqrt{1-\Sigma_j\delta p_j}$, remaining in the same manifold. Here, $c_j$ are the collapse operators $\sqrt{\kappa_1}a,\sqrt{\kappa_2}b,\sqrt{P}\sigma_+,\sqrt{\gamma}\sigma_-$. When a quantum jump $j$ occurs, the system makes a transition to the state $\ket{\psi(t+\delta t)}=c_j\ket{\psi(t)}/\sqrt{\delta p_j}$. The four possible quantum jumps correspond respectively to the emission of a photon from the cavity mode at $\omega_0$, emission of a photon from the cavity mode at $\omega_0/2$, atom being pumped to the excited state, and the decay of the emitter to the ground state through pathways other emission into the cavity modes. Hence, a single Monte-Carlo trajectory allows us to visualize the photon emission processes\cite{jumpform,click1,wiseman1996quantum}. \\

\begin{figure*}[hbt!]
\includegraphics[width=150mm,scale=0.8]{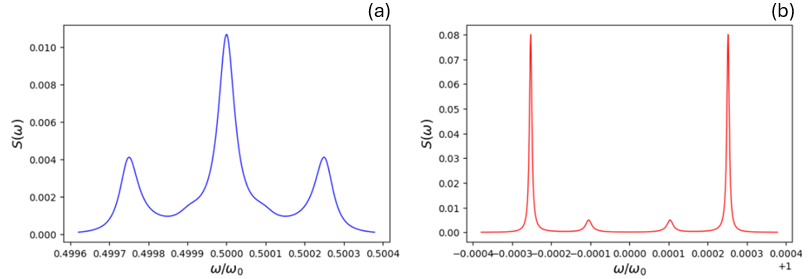}
\caption{\label{fig:Spectrum}  Subfigure (a) on the left illustrates the spectrum of the $\omega_0/2$ mode, while subfigure (b) depicts the spectrum of the $\omega_0$ mode. Here, $\kappa_1=0.02g_1$, $\gamma=0.016g_1$, $\kappa_2=0.2g_1$, $P=0.005g_1$.} 
\end{figure*}
\begin{figure*}[hbt!]
\includegraphics[width=100mm,scale=1]{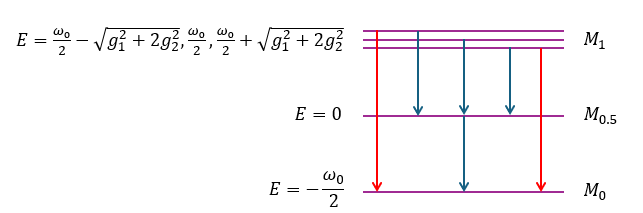}
\caption{\label{fig:SpectrumExpl} The energy levels and transitions of the system are illustrated in the diagram. The purple lines represent the five energy levels of the system. The red arrows indicate the transitions that lead to the two prominent peaks in the spectrum for the $\omega_0$ mode. Meanwhile, the blue arrows correspond to the transitions associated with the peaks in the spectrum of the $\omega_0/2$ mode.}
\end{figure*}

Figure \ref{MC2}(a) and (b) show a single trajectory of the system for two different cavity designs. The y-axis represents the expectation values of various populations, plotted as a function of time $g_1t$. The red curve corresponds to the mean excited state population $\langle \ket{e}\bra{e} \rangle$ and blue curve corresponds to the mean photon count $\langle b^{\dagger}b\rangle$. Fig.\ref{MC2}(a) is the trajectory of a system with $\kappa_1=0.2g_1$ and a lower efficiency of $8.16 \%$. Fig. \ref{MC1}(b) depicts the evolution of another system with lower $\kappa_1=0.02g_1$ and thus a higher efficiency of $35.41\%$. The remaining parameters are taken to be identical in the two cases and equal to the default value as specified before. The system starts out from the initial state of $\ket{e,0,0}$. Both figures show two visually distinct regions: a densely shaded region where the populations oscillate rapidly, interrupted by regions where all populations are zero. The system is in the manifold $M_0$, i.e., state $\ket{g,0,0}$ in the region with all zero populations. The system remains in $M_0$ until $\sqrt{P}\sigma_+$ causes a jump to the manifold $M_1$, which corresponds to the densely shaded region. The zoomed in view of the system dynamics in two such densely shaded regions are showed in Figures \ref{MC2}(c) and \ref{MC2}(d). These show fast Rabi oscillations with the population oscillating primarily between the states $\ket{e,0,0}$ and $\ket{g,1,0}$ at a frequency $g_1/(2\pi)$. The population in the state $\ket{g,0,2}$ remains quite small as $g_2<g_1$. Thus, the densely shaded regions correspond to the coherent evolution of the system within the manifold $M_{1}$ under the influence of $H_{e}$. The one- and two-photon Rabi oscillations in this manifold continue until a quantum jump occurs, which takes the system to a different manifold. Since these simulations were conducted in the low $P$ regime, we observe only the dissipative quantum jumps that take the system to one of the lower manifolds $M_{0.5}$ and $M_0$. \\

Two types of dissipative quantum jumps are observed: the first type, where the system jumps from $M_1$ to $M_0$ accompanied by a vertical blue spike, and the other where the system jumps to $M_0$ without the blue spike. These are further analyzed in Figure \ref{MC2}(c) and (d). The second type, depicted in the subfigure (c), occurs when the system transitions directly from $M_{1}$ to \(M_{0}\). This is indicated by all populations going down to zero simultaneously when this type of jump occurs. These jumps are caused by the collapse operators $\sqrt{\kappa_1}a$ or $\sqrt{\gamma}\sigma_-$, resulting in one-photon emission or spontaneous emission into non-resonant modes. The other type of quantum jump is depicted in the subfigure (d). Here, $\langle b^{\dagger}b\rangle$ first shoots up to unit probability, whereas $\langle \ket{e}\bra{e} \rangle$ goes down to zero. The unit value of $\langle b^{\dagger}b\rangle$ indicates the system transitioning from $M_1$ to $M_{0.5}$ i.e., $\ket{g,0,1}$, where the emitter is in the ground state and the number of photons in the cavity is one. This jump is caused by $\sqrt{\kappa_2}b$ and results in the emission of the first photon of the two-photon state from the cavity. The emission of the second $\omega_0/2$ photon occurs in quick succession, and the system jumps down to $M_0$. \\

Thus, the two-photon emission is a fast cascade process where the state of the system jumps from coherent evolution in $M_{1}$ to $M_{0.5}$, thereby releasing the first photon, followed quickly by another jump to $M_{0}$ releasing the second photon.\\

After the system decays to $M_0$ via these two types of jumps, after some time it is once again pumped up to $M_{1}$ by the action of the incoherent pump.

In the high efficiency case, the cascade jump process during a specific time period, i.e., the TPE, occurs more frequently than in the low efficiency case. This is illustrated in Figure \ref{MC2}(a) and (b), where the subfigure (a) depicts the high-efficiency case, showing eight instances of TPE, whereas the subfigure (b) depicts the low-efficiency case, showing only one instance of TPE during the period of time for which the simulation was conducted. The connection between the efficiency and the frequency or probability of the cascade jump can be illustrated by analyzing the no-jump case. The state of the system $\ket{\psi(t)}$ evolving under $H_e$ in the manifold $M_1$ can be expressed as: 

\begin{align}
    \frac{d\ket{\psi(t)}}{dt} &= -iH_e\ket{\psi(t)}
\end{align}
Here
\begin{align}
    \ket{\psi(t)} &=c_1(t)\ket{e,0,0} +c_2(t)\ket{g,1,0} + c_3(t)\ket{g,0,2}
\end{align}

The instantaneous TPE rate at time $t$ is $T(t)=\kappa_2\bra{\psi(t)}b^{\dagger}b\ket{\psi(t)} = 2\kappa_2|c_3(t)|^2$. The time-integration of TPE rate quantifies the occurrence probability $P_T$ of the quantum jump causing the two-photon emission. It is given by:
\begin{align}
    P_T=\int_{0}^\infty T(t)dt \approx \frac{4\kappa_2g_2^2}{(\gamma+\kappa_1)(\kappa_2^2+g_1^2)}=\frac{\eta}{100}
\end{align}

The details of the calculation are given in Appendix \ref{sec: ManifoldProb}. Thus, the efficiency of two-photon emission corresponds to the probability of the cascade quantum jump process occurring.\\

\subsection{Spectrum Of Emission}

The spectrum of the cavity output field for the two modes can be calculated from the cavity field correlations (obtained from the master equation) using the formulae:
\begin{align}
    S_a(\omega)&=\int_{-\infty}^{\infty}\langle a^{\dagger}(\tau)a(0) \rangle e^{-i \omega \tau} d\tau \\
    S_b(\omega)&=\int_{-\infty}^{\infty}\langle b^{\dagger}(\tau)b(0) \rangle e^{-i \omega \tau} d\tau
\end{align}
Here, $S_a(\omega)$ and $S_b(\omega)$ are the cavity emission spectra for the $\omega_0$ and $\omega_0/2$ modes respectively. These are plotted as a function of frequency $\omega$ in Figure \ref{fig:Spectrum}. $S_b(\omega)$, plotted in Figure \ref{fig:Spectrum}(a), is symmetric around $\omega_0/2$, with a large peak centered at $\omega_0/2$, and two smaller peaks on either side. $S_b(\omega)$, plotted in Figure \ref{fig:Spectrum}(b), is also symmetric with two large peaks on either side of $\omega_0$, and two  smaller peaks closer to $\omega=\omega_0$. \\

The various peaks in the spectra correspond to the different transitions between the manifolds of the system \cite{scully1997quantum,qjump}. Taking into account only the lowest three manifolds, the diagonalization of the system Hamiltonian $H$(Eq.\ref{Ham}) results in 5 energy levels of the system, which are depicted in Figure \ref{fig:SpectrumExpl} by the solid horizontal lines. The lowest two levels are the states $\ket{g,0,0}$ and $\ket{g,0,1}$ with energies $-\omega_0/2$ and $0$ respectively. The higher three states belong to the third manifold $M_{1}$, with energies $E_-=\omega_0/2-\sqrt{g_1^2+2g_2^2},E_0=\omega_0/2$ and $E_+=\omega_0/2 + \sqrt{g_1^2+2g_2^2}$. The allowed transitions between these levels, indicated by vertical arrows, give rise to the peaks observed in Figure \ref{fig:Spectrum}. The transitions contributing to the $S_b(\omega)$ spectrum are depicted by four blue arrows whereas the transitions corresponding to the peaks in the $S_a(\omega)$ spectrum are depicted by two red arrows. The transitions from the $E_+$ and $E_-$ energy levels of $M_{1}$ to $\ket{g,0,1}$ explain the right and left peaks at $\omega=\omega_0/2 \pm \sqrt{g_1^2+2g_2^2}$ in the $S_b(\omega)$ spectrum. The remaining two transitions shown in blue contribute to the central peak at $\omega_0/2$ in $S_b(\omega)$. This is because the energy separation between the states $E_0$ and $\ket{g,0,1}$, and that between the states $\ket{g,0,1}$ and $\ket{g,0,0}$ are identical. This makes the central peak of $S_b(\omega)$ higher than the two side peaks. The red arrows show the transitions associated with one-photon emission. The transition from the upper energy level of $M_1$ at $E_+$ to $\ket{g,0,0}$ gives rise to the large peak on the right of $\omega=\omega_0$. The transition from the lower energy level of $M_1$ at $E_-$ to $\ket{g,0,0}$ results in the large peak to the left of $\omega=\omega_0$. The eigenstate $\ket{\psi_0}$ corresponding to the energy level of $M_{1}$ at $\omega_0/2$ is approximately $c_1\ket{g,0,2} + c_2\ket{g,1,0}\approx \ket{g,0,2}$ as $c_2/c_1=-\sqrt{2}g_2/g_1$. Thus, this state cannot jump to a lower manifold via one-photon emission. Therefore, the central peak for the one-photon emission spectrum is negligible and cannot be seen. The two small peaks close to $\omega=\omega_0$ are due to transitions involving higher order manifolds.\\

The location of the peaks predicted by the above analysis agree with the numerical results. For the chosen parameters, $\sqrt{g_1^2+2g_2^2}\approx g_1 \approx \omega_0/4000$, resulting in peaks at $\omega_0/2 \pm 0.00025$, $\omega_0/2$ and $\omega_0 \pm 0.00025$. This is exactly what is seen in Figure \ref{fig:Spectrum}.

\section{Conclusion}

We have done a comprehensive theoretical study of an emitter in a doubly resonant cavity for efficient photon-pair generation. Using a combined analytical–numerical framework, we have identified an optimal cavity design for maximizing the efficiency of photon-pair generation. The optimal cavity design requires simultaneous lowering of the out-coupling rate at one-photon emission frequency to below that of the atomic loss rate ($\kappa_1<\gamma$) as well as equalizing the out-coupling rate of the cavity mode at photon-pair emission frequency to match the one-photon vacuum Rabi frequency ($\kappa_2\approx g_1$). With experimentally demonstrated parameters, we achieve a two-photon generation efficiency of approximately $35\%$, which is several times higher than of the parametric down conversion-based methods. The efficiency can be further increased to around $50\%$ by further reducing $\kappa_1$ or increasing the quality-factor $Q_1$ to $10^6$. Furthermore, the highest efficiency is only attainable at low pump rates. The maximum two-photon emission rate occurs at relatively higher $P$, with an emission rate of approximately $3\times10^5$ photon pairs per second at an efficiency of around $12\%$, which is still higher than that of SPDC. \\

The cavity field of the $\omega_0/2$ mode exhibits super-Poissonian photon statistics and the emitted two-photons are highly bunched with a $g^2(0)$ value of several hundred. The two-photon emission mechanism involves a rapid cascade process of quantum jumps, where the second photon gets emitted from the cavity shortly after the first one. The cavity emission spectrum exhibits three closely spaced distinct peaks at and around the two-photon emission frequency $\omega_0/2$. \\


   
\appendix 
\onecolumngrid

\section{\label{sec:derivation}Derivation of the Hamiltonian}

\begin{figure*}[hbt!]
\includegraphics[width=140mm,scale=0.8]{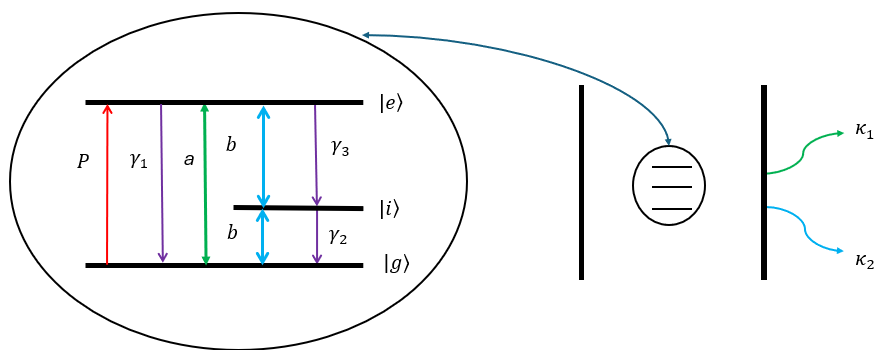}
\caption{\label{fig:3level} The atom consists of levels denoted by $\ket{g}$,$\ket{e}$ and $\ket{i}$, interacting with the $a$ and $b$ modes of the cavity, denoted by bidirectional arrows in blue and green. The dashed line represents the zero energy level, in-between $\ket{g}$ and $\ket{e}$ which are $\omega_0/2$ above and below it. The unidirectional purple arrows denote the incoherent decay processes.}
\end{figure*}

The Hamiltonian of the system \( H \) (Eq. \ref{Ham}), can be realized as an effective model of a three-level atom that interacts with the two modes of a doubly resonant cavity. The system is illustrated in Figure \ref{fig:3level}. In addition to the two atomic levels $\ket{g}$ and $\ket{e}$, there is now a third intermediate level, $\ket{i}$, with $\omega_0/2+\delta$ above $\ket{g}$ (or $\omega_0/2-\delta$ below $\ket{e}$). The interactions between the atom and the fields are modeled using the standard Jaynes-Cummings model. In this model, the $\omega_0$ mode couples the atomic states $\ket{g}$ and $\ket{e}$, whereas the $\omega_0/2$ mode drives transitions between $\ket{g}$ and $\ket{i}$ and between $\ket{i}$ and $\ket{e}$. As discussed below, we also require $\delta$ to be sufficiently larger than all the coherent coupling strengths $g_1,g_3,g_4$. Such a system can be implemented in double quantum dots \cite{DQD} or artificial atoms such as qutrits formed by superconducting circuits\cite{cyclic3,cyclic3_2,cyclic3_3}. The bi-directional arrows in Figure \ref{fig:3level} represent coherent transitions. The unidirectional arrows represent the incoherent processes including the atomic decay and pumping at rates given by $\gamma_1$ and $P$ respectively. The outcoupling rates of the $\omega_0$ and $\omega_0/2$ cavity modes are denoted by $\kappa_1$ and $\kappa_2$. The Hamiltonian of the three-level system is:

\begin{widetext}
    \begin{align}
        H_3=\frac{1}{2}\omega_0(\ket{e}\bra{e}-\ket{g}\bra{g})+\delta \ket{i}\bra{i}+ \frac{1}{2}\omega_0 b^{\dagger}b +\omega_0 a^{\dagger}a + g_1(a^\dagger\ket{g}\bra{e}+a\ket{e}\bra{g})+g_3(b^{\dagger}\ket{g}\bra{i}+b\ket{i}\bra{g})+g_4(b^{\dagger}\ket{i}\bra{e}+b\ket{e}\bra{i})  
    \end{align}
\end{widetext}

\twocolumngrid

The Hamiltonian $H$ (eq. \ref{Ham}), can be obtained by performing a unitary transformation on \( H_3 \), resulting in \( H_3' = e^S H e^{-S} \) with
\begin{align} \label{S}
    S=\left(\frac{g_4}{\delta}b^{\dagger}\ket{i}\bra{e}-\frac{g_3}{\delta}b^{\dagger}\ket{g}\bra{i}\right) -H.c.
\end{align}

Here, H.c. stands for Hermitian conjugate. Using the Baker–Campbell–Hausdorff formula, the transformed Hamiltonian can be expressed as a series with terms involving different powers of $S$. 
\begin{align}
    H_3'&=e^SH_3e^{-S} \notag\\
    &=H_3 + [S,H] + \frac{1}{2!}[S,[S,H]]+ h.o.t
\end{align}
Here, h.o.t refers to the higher order terms with coefficients proportional to higher power of $g_i/\delta$. As magnitude of $\delta$ is chosen to be sufficiently larger than $g_1,g_3,g_4$, the terms involving second or higher powers of $S$, and thus $g_ig_j/\delta$ ($i,j=1,3,4$) can be ignored in the transformed Hamiltonian. This results in

\begin{figure*}[hbt!]
\includegraphics[width=150mm,scale=1]{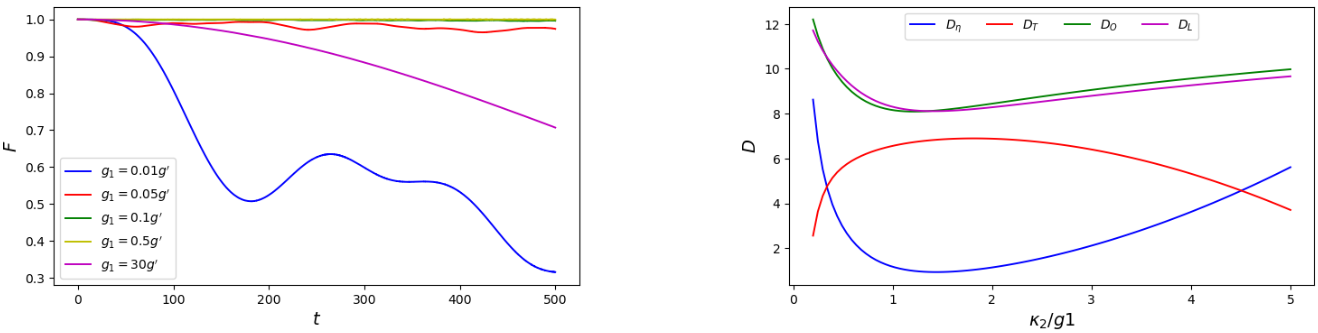}
\caption{\label{fig:Dg2} The Sub-figure on the left illustrates the fidelity between the states evolving under the influence of $H_3$ and $H_m$, for different values of $g_1$, as a function of time normalised by $g'$. The magnitude of $\delta=100g'$. The Sub-figure on the right shows the deviation in efficiency $D_{\eta}$ and deviation in $\langle \ket{g}\bra{g} \rangle$ as a function of $\kappa_2/g1$. The other parameters are $P=0.005g_1,\kappa_1=0.02g_1,\gamma=0.016g_1,g_1=0.1g'$ }. 
\end{figure*}

\begin{equation}
    H_3'=H_m+H_a
\end{equation}
where 
\begin{widetext}
    \begin{align}
        H_m=\frac{1}{2}\omega_0(\ket{e}\bra{e}-\ket{g}\bra{g})+ \frac{1}{2}\omega_0 b^{\dagger}b +\omega_0 a^{\dagger}a + g_1(a^\dagger\ket{g}\bra{e}+a\ket{e}\bra{g})+\frac{g_3^2+g_4^2-4g_3g_4}{2\delta}(b^{\dagger2}\ket{g}\bra{e}+b^2\ket{e}\bra{g})
    \end{align}
\end{widetext}

and
\newpage
\begin{widetext}
    \begin{align}
        H_a&=\delta \ket{i}\bra{i} + \frac{g_1g_4}{\delta}(a^{\dagger}b\ket{g}\bra{i})+H.c.)+\frac{g_1g_3}{\delta}(a^{\dagger}b\ket{i}\bra{e}+H.c.) + \frac{2g_4^2-g_3g_4}{\delta}(b^{\dagger}b\ket{i}\bra{i} - bb^{\dagger}\ket{e}\bra{e}) + \frac{g_3g_4-2g_3^2}{\delta}(b^{\dagger}b\ket{g}\bra{g}
         - bb^{\dagger}\ket{i}\bra{i}) +\notag \\
    \end{align}
\end{widetext}

Note that since $\delta>>g_3,g_4$, the unitary transformed dressed basis is approximately the same as the original basis\cite{detuning1,detuning2}. Thus, in the large $\delta$ limit, the evolution of the state in the original basis occurs due to our effective Hamiltonian $H_3'$.\\

The term \( H_m \) is equivalent to the two-level Hamiltonian \( H \), with the relation \( g_2=(g_3^2+g_4^2-4g_3g_4)/2\delta\). \( H_a \) represents the extra terms in \( H_3' \) that cause the system's dynamics to deviate from that of the two-level model. Therefore, $H_m$ serves as a valid effective Hamiltonian only in regions of the parameter space where the contributions from \( H_a \) can be neglected. To identify this region, we carry out numerical simulations of the time evolution of the system due to $H_3$ and $H_m$ and evaluate the fidelities for different parameter values. We also compare the steady-state expectation values of operators predicted by both the models. For the two-level system, we use the same Lindbladian as given by Eq. $\ref{Lindblad2}$, whereas for the 3-level system, we analyze the steady-state of the Master equation:  

\begin{widetext} \label{L_3}
    \begin{align} \label{L_3}
        \frac{d\rho}{dt}=-i[H_3,\rho]+\frac{\kappa_1}{2}\mathcal{L}_{a}\rho + \frac{\kappa_2}{2}\mathcal{L}_{b}\rho +\frac{P}{2}\mathcal{L}_{\ket{e}\bra{g}}\rho+\frac{\gamma}{2}\mathcal{L}_{\ket{g}\bra{e}}\rho
    \end{align}
\end{widetext}

We ignore the atomic decay between levels $\ket{e}$ and $\ket{i}$, as well as between $\ket{i}$ and $\ket{g}$, because for large detunings, the density of states of the resonator at $\omega = \omega_0/2 \pm \delta$ is negligible. Therefore, the spontaneous emission rates at these frequencies are suppressed and can be neglected i.e., $\gamma_2=\gamma_3=0$. For the sake of convenience, we take $g_3=g_4=g'$ for conducting the simulations. The value of \(g'\) is set to 1, and the value of \(\omega_0\) is taken from Ref. \cite{cyclic_params}. The ratio of \(g'/\omega_0\) is set as \(1/400\),because $\omega_0$ is twice the transition frequency between adjacent levels as mentioned in Ref.\cite{cyclic_params}. The magnitude of \(\delta\) is taken to be \(\delta = 100g'\), which satisfies the $\delta>>g'$ condition.\\

The fidelity between the states evolving under $H_3$ and $H_m$, initially prepared in $\ket{e,0,0}$, are depicted in the subfigure on the left in Figure \ref{fig:Dg2}. The Fidelity is low for low values of $g_1/g'$ and increases as $g_1$ increases. For $g_1\geq 0.1g'$, the Fidelity is approximately 1. The reason for this is that at low $g_1$, the energy shift term in $H_a$ i.e., $[(2g_4^2-g_3g_4)/\delta](b^{\dagger}b\ket{i}\bra{i} - bb^{\dagger}\ket{e}\bra{e}) + [(g_3g_4-2g_3^2)/\delta](b^{\dagger}b\ket{g}\bra{g}-bb^{\dagger}\ket{i}\bra{i})$ causes significant deviations in the dynamics.  As $g_1$ increases, the effect of the energy shift terms decreases, and above $g'=0.1g_1$, the energy shift term acts as a perturbation, resulting in the Fidelity rising to almost 1, indicating that $H_m$ is an excellent effective Hamiltonian of $H_3$. The transition term i.e., $[g_1g_4/\delta](a^{\dagger}b\ket{g}\bra{i})+H.c.)+[g_1g_3/\delta (a^{\dagger}b\ket{i}\bra{e}+H.c.)$, which causes two-photon transitions between $\ket{i}$ and $\ket{g},\ket{e}$ is negligible at low $g_1$. However, at still higher values of $g'$, the fidelity once again decreases because the effect of the transition term becomes prominent, causing deviation between the time-evolution due to $H_3$ and $H_m$. Thus, for $g_1\geq0.1g'$, our model Hamiltonian $H_m$ is a valid effective Hamiltonian. \\

The maximum efficiency occurs for the highest value of $g_2/g_1=-g'^2/(\delta g_1)$. Thus, we set the value of $g_1=0.1g'$. To obtain $H_m$, we used the unitary transformation $U=e^S$ (Eq. \ref{S}) on $H_3$ and neglected $H_a$. Similarly, to obtain the corresponding Master equation for $H_m$, we perform a unitary transform on the entire equation \ref{L_3} and we neglect all terms which are not present in Eq. \ref{Lindblad2}. This causes deviations in the steady-state behaviour as predicted by Eqs. \ref{L_3} and \ref{Lindblad2}. To measure the deviations and check the validity of our model master equation, we plot the Mean Absolute Percentage Deviation (MAPD) for steady-state expectation values of various operators. The  MAPD of an operator $X$ i.e., $D_X$ is defined as:

\begin{equation}
    D_X=\frac{X_{3}-X_{2}}{X_{2}}\times 100
\end{equation}

Here $X_{3}$ and $X_{2}$ refer to the steady state expectation values of an operator $X$ as predicted by the simulation of the full three-level system (Eq.\ref{L_3}) and the two-level model respectively (Eq. \ref{Lindblad2}). We analyse the steady-state values of $\eta$ (Eq. \ref{eff_expr}) and $T,O,L$ (Eqs.\ref{TPE},\ref{OPE},\ref{loss}), denoted by $D_{\eta}$ and $D_T,D_O,D_L$ respectively. Note that using $\eta,T,O,L$ we can also find out the values of all other operators, as mentioned in Appendix \ref{sec:SSE}. Therefore, if the MAPD for these 4 quantities is low, then the MAPD for all other expectation values of operators is also low. The unitary transformation $U=e^S$ (Eq.\ref{S}) does not affect $(\kappa_1/2)\mathcal{L}_a \rho,$ due to the absence of $a$ in $S$, but it affects the other three namely $(\kappa_2/2)\mathcal{L}_b \rho,(\gamma/2)\mathcal{L}_{\sigma_-} \rho,(P/2)\mathcal{L}_{\sigma_+}\rho$. Out of these three, the only variable parameter is $\kappa_2$. This indicates that low MAPD values of $\eta,T,O,L$ across the entire range of $\kappa_2$ considered in Section \ref{sec:sectionSSR} demonstrate that our model is valid for the parameter range selected in our analysis. Thus, we plot $D_{\eta}$ and $D_T,D_O,D_L$ as a function of $\kappa_2$, at $P=0.005g_1,\kappa_1=0.02g_1,\gamma=0.016g_1$, which are the same as the parameters used in Section \ref{sec:sectionSSR}. We see that the deviations remain of the order of $10\%$ in all cases. The maximum deviation is in $O,L$ at low $\kappa_2$. In contrast, $D_{\eta}$ and $D_T$ remain under $8\%$ in all cases. At $\kappa_2=g_1$, which is the point of optimal two-photon emission rate and efficiency, all deviations are less than $8\%$ and the deviation in efficiency is less than $2\%$. \\

Thus, our system model (Eq. \ref{Lindblad2}) as described in Section \ref{sec:section2} is a valid effective model of the three-level system (Eq. \ref{L_3}) for $g_1 = 0.1g'$ and $\delta=100g'$, corresponding to $g_2=0.1g_1$.

\section{\label{sec:SSE}Steady State Equations} 

The Hilbert space of the system is infinite-dimensional, which results in an infinite number of equations for the steady-state expectation values of various operators. However, by making the approximation that the state of the system remains confined to the first three manifolds, we arrive at a finite set of equations. Let the dimension of the Hilbert Space be $n_1\times n_2 \times n_3$ where $n_1$ refers to the dimensions of the Hilbert space of the atom, $n_2$ is the dimension of the $\omega_0$ mode, and $n_3$ is the dimension of the $\omega_0/2$ mode. In this approximation, the highest Fock states of the $\omega_0$ and $\omega_0/2$ modes are 1 and 2 respectively. The atom is considered to be a two-level system. Therefore, the dimension of the Hilbert Space is $2 \times 2 \times 3$ i.e., a twelve dimensional Hilbert Space. The steady-state density matrix has dimensions of $12 \times 12$. Calculation of the steady-state expectation values of operators results in 18 equations in terms of 18 steady state expectation values, while the expectation values of all other operators are zero. These equations are listed as follows:
\begin{widetext}
    \begin{align}
        \kappa_1 \langle a^{\dagger}a \rangle =& -ig_1 \langle a^{\dagger}\sigma_- - a\sigma_+ \rangle  \\
        \frac{\kappa_2}{2} \langle b^{\dagger}b \rangle =& -ig_2 \langle b^{\dagger 2}\sigma_- - b^2\sigma_+ \rangle \\
        P \langle \ket{g}\bra{g} \rangle =& \kappa_1 \langle a^{\dagger}a \rangle + \frac{\kappa_2}{2} \langle b^{\dagger}b \rangle + \gamma \langle \ket{e}\bra{e} \rangle \\
        (\frac{P+\gamma}{2}+\kappa_2)\langle b^{\dagger 2}\sigma_- \rangle =& -i\{g_1[\langle 2 a b^{\dagger 2}\ket{g}\bra{g}- a b^{\dagger 2} \rangle] + g_2 [\langle (2b^{\dagger 2}b^2+4b^{\dagger}b+2)\ket{g}\bra{g} -(b^{\dagger 2}b^2+4b^{\dagger}b+2)\rangle]\}\\
        (\frac{\kappa_1+P+\gamma}{2})\langle a^{\dagger}\sigma_- \rangle =& i\{-g_2[ \langle 2a^{\dagger}b^2\ket{g}\bra{g}-a^{\dagger}b^2 \rangle ]-g_1[\langle 2 a^{\dagger}a\ket{g}\bra{g}-a^{\dagger}a -\ket{e}\bra{e}\rangle]\} \\
        (\kappa_2+\frac{\kappa_1}{2} + P + \gamma)\langle a^{\dagger}b^2\ket{g}\bra{g} \rangle =& -i\{-g_1[\langle a^{\dagger}ab^2\sigma_+ + b^2\sigma_+\rangle]+g_2[\langle(b^{\dagger 2}b^2+4b^{\dagger}b+2)a^{\dagger}\sigma_- \rangle] + \gamma\} \\
        (\kappa_2+\frac{\kappa_1}{2})\langle a^{\dagger}b^2 \rangle=& -i\{2g_2[\langle a^{\dagger}\sigma_-b^{\dagger}b + a^{\dagger}\sigma_-]-g_1 \langle b^2\sigma_+ \rangle \} \\
        \langle b^{\dagger 2}b^2 \rangle =& \frac{1}{2} \langle b^{\dagger}b \rangle\\
        (\kappa_1+\kappa_2+\frac{P+\gamma}{2})\langle a^{\dagger}ab^2\sigma_+ \rangle =& i\{g_1\langle b^2[a^{\dagger}\ket{g}\bra{g}-a^{\dagger}] \rangle+g_2\langle a^{\dagger}a[(2b^{\dagger 2}b^2+4b^{\dagger}b+2)\ket{g}\bra{g}- (b^{\dagger 2}b^2+4b^{\dagger}b+2)] \rangle \} \\
        (\kappa_1+P+\gamma)\langle a^{\dagger}a\ket{g}\bra{g} \rangle =&-i\{-g_1\langle a\sigma_+ \rangle + g_2 \langle a^{\dagger}a[b^{\dagger 2}\sigma_- - b^2\sigma_+] \rangle + \gamma\}\\
        (\kappa_2+\frac{\kappa_1+P+\gamma}{2})\langle b^{\dagger}b a^{\dagger} \sigma_- \rangle =& -i\{ g_1 \langle b^{\dagger}b[(2 a^{\dagger}a+1)\ket{g}\bra{g}-(a^{\dagger}a+1)]\rangle +2g_2 \langle a^{\dagger}[b^2\ket{g}\bra{g}-b^2]\rangle \}\\
        (2\kappa_2+\frac{\kappa_1+P+\gamma}{2})\langle b^{\dagger 2}b^2 a^{\dagger} \sigma_- \rangle =& -i\{ g_1 \langle b^{\dagger 2}b^2[(2 a^{\dagger}a+1)\ket{g}\bra{g}-(a^{\dagger}a+1)]\rangle +2g_2 \langle a^{\dagger}[b^2\ket{g}\bra{g}-b^2]\rangle \}\\
        (\kappa_1+\kappa_2+P+\gamma)\langle a^{\dagger}a b^{\dagger}b \ket{g}\bra{g} \rangle =& i\{g_1 \langle b^{\dagger}b a \sigma_+ \rangle + 2g_2 \langle a^{\dagger}a b^2 \sigma_+ \rangle +\gamma \} \\
        (\kappa_1+2 \kappa_2+P+\gamma)\langle a^{\dagger}a b^{\dagger 2}b^2 \ket{g}\bra{g} \rangle =& i\{g_1 \langle b^{\dagger 2}b^2 a \sigma_+ \rangle + 2g_2 \langle a^{\dagger}a b^2 \sigma_+ \rangle +\gamma \} \\
        (\kappa_2+P+\gamma)\langle b^{\dagger}b \ket{g}\bra{g} \rangle =& -i\{g_1 \langle b^{\dagger}b[a^{\dagger}\sigma_- - a \sigma_+] \rangle -2 g_2 \langle b^2 \sigma_+ \rangle + \gamma \} \\
        (2 \kappa_2+P+\gamma)\langle b^{\dagger 2}b^2 \ket{g}\bra{g} \rangle =& -i\{g_1 \langle b^{\dagger 2} b^2 [a^{\dagger}\sigma_- - a \sigma_+] \rangle -2 g_2 \langle b^2 \sigma_+ \rangle + \gamma \} \\
        (\kappa_1+\kappa_2) \langle a^{\dagger}a b^{\dagger}b \rangle =& -i \{ g_1 \langle b^{\dagger}b [a^{\dagger}\sigma_- - a \sigma_+]\rangle + 2 g_2 \langle a^{\dagger}a[b^{\dagger 2} \sigma_- - b^2 \sigma_+] \rangle \} \\
        (\kappa_1+2 \kappa_2) \langle a^{\dagger}a b^{\dagger 2} b^2 \rangle =& -i \{ g_1 \langle b^{\dagger 2} b^2 [a^{\dagger}\sigma_- - a \sigma_+]\rangle + 2 g_2 \langle a^{\dagger}a[b^{\dagger 2} \sigma_- - b^2 \sigma_+] \rangle \} 
    \end{align}
\end{widetext}
These can be simplified greatly using the approximation that the trace goes over only the 5 basis states of the three manifolds,as mentioned in subsection \ref{sec:ManifoldApprox}. For example:

\begin{eqnarray}
    \langle a^{\dagger}b^2 \rangle &=& \sum_{i=1}^n\bra{i} a^{\dagger}b^2 \rho  \ket{i} \nonumber \\ &=& \sqrt{2} \bra{g,0,2} \rho \ket{g,1,0} \\
    \langle a^{\dagger}b^2 \ket{g}\bra{g} \rangle &=& \sum_{i=1}^n\bra{i} a^{\dagger}b^2 \ket{g}\bra{g}  \rho \ket{i} \nonumber \\
    &=& \sqrt{2} \bra{g,0,2} \rho \ket{g,1,0}
\end{eqnarray}

Hence, $\langle a^{\dagger}b^2 \rangle=\langle a^{\dagger}b^2 \ket{g}\bra{g} \rangle$. The other relations between various operators are listed as follows: \\
\begin{align}
    \langle a^{\dagger}a  b^{\dagger}b \rangle &= \langle a^{\dagger}a  b^{\dagger}b \ket{g}\bra{g} \rangle=0 \\
    \langle a^{\dagger}a  b^{\dagger 2}b^2 \rangle &= \langle a^{\dagger}a  b^{\dagger 2}b^2 \ket{g}\bra{g} \rangle=0 \\
    \langle a^{\dagger}a \rangle &= \langle a^{\dagger}a \ket{g}\bra{g} \rangle \\
    \langle b^{\dagger}b \rangle &= \langle b^{\dagger}b \ket{g}\bra{g} \rangle \\
    \langle b^{\dagger 2}b^2 \rangle &= \langle b^{\dagger 2}b^2 \ket{g}\bra{g} \rangle
\end{align}

This leads to 4 equations in 4 variables:
\begin{widetext}
    \begin{align}
        (\frac{\kappa_2(\kappa_2+P/2+\gamma/2)}{2g_2} +g_2) \langle b^{\dagger}b \rangle &= 4g_2 \langle \ket{e}\bra{e} \rangle - g_1 \langle a^{\dagger}b^2 +a b^{\dagger 2} \rangle \\
        (\frac{\kappa_1(\kappa_1+P+\gamma)}{2g_1} + 2g_1) \langle a^{\dagger}a \rangle &= 2g_1 \langle \ket{e}\bra{e} \rangle -   g_2 \langle a^{\dagger}b^2 +a b^{\dagger 2} \rangle \\
        (\kappa_2 + \frac{\kappa_1}{2}) \langle a^{\dagger}b^2 +a b^{\dagger 2} \rangle &= \frac{2 \kappa_1 g_2}{g_1} \langle a^{\dagger}a \rangle + \frac{\kappa_2 g_1}{2 g_2} \langle b^{\dagger}b \rangle \\
        P \langle \ket{g}\bra{g} \rangle &=\kappa_1 \langle a^{\dagger}a \rangle + \frac{\kappa_2}{2} \langle b^{\dagger}b \rangle + \gamma \langle \ket{e}\bra{e} \rangle 
    \end{align}  
\end{widetext}

Solving these equations results in the closed-form solutions as given in subsection \ref{sec:ManifoldApprox}, and these closed-form solutions also enable the calculation of all the other 18 non-zero steady state expectation values.

\section{\label{sec: ManifoldProb}Probability of Two-Photon Quantum Jump} 

The equations for the time evolution of $c_1(t),c_2(t),c_3(t)$ are: 

\begin{align}
    i\frac{\partial c_1}{\partial t} &= \frac{\omega_0-i\gamma}{2} c_1 + g_1c_2 + \sqrt{2}g_2c_3 \label{eq1}\\
    i\frac{\partial c_2}{\partial t} &= g_1c_1 + \frac{\omega_0-iP-i\kappa_1}{2} c_2 \label{eq2} \\
    i\frac{\partial c_3}{\partial t} &= \sqrt{2}g_2c_1 + (\frac{\omega_0-iP}{2}-i\kappa_2) c_3  \label{eq3}
\end{align}

Since $g_2<<g_1$, we ignore $g_2$ in Eq. \ref{eq1}. This results in the following system of equations for $c_1(t)$ and $c_2(t)$: 

\begin{align}
    i\frac{\partial c_1}{\partial t} &= \frac{\omega_0-i\gamma}{2} c_1 + g_1c_2 + \label{eqa1}\\
    i\frac{\partial c_2}{\partial t} &= g_1c_1 + \frac{\omega_0-iP-i\kappa_1}{2} c_2 \label{eqa2} 
\end{align}

Solving these with the initial conditions of $c_1(0)=1,c_2(0)=0$ gives:

\begin{align}
    c_1(t)&\approx\frac{e^{\lambda_1t}+e^{\lambda_2t}}{2} \\
    c_2(t)&\approx\frac{e^{\lambda_1t} - e^{\lambda_2t}}{2}
\end{align}

where 

\begin{align}
    \lambda_1&=-i(\frac{\omega_0}{2} + g_1) - \frac{\gamma+P+\kappa_1}{4} \\
    \lambda_2&=-i(\frac{\omega_0}{2} - g_1) - \frac{\gamma+P+\kappa_1}{4} 
\end{align}

Let $c_3'(t)=c_3e^{i(\omega_0/2)}t$. Since $e^{i(\omega_0/2)t}$ is a phase factor, $|c_3(t)|^2=|c_3'(t)|^2$. We find $c_3'(t)$, with the initial condition $c_3(0)=0$ as:

\begin{align}
    c_3'(t)=\frac{-ig_2}{\sqrt{2}}(\frac{e^{(\lambda_1+i\omega_0/2)t}}{\lambda_1+\theta} + \frac{e^{(\lambda_2+i\omega_0/2)t}}{\lambda_2+\theta}-\frac{e^{-\theta t}}{\lambda_1+\theta} -\frac{e^{-\theta t}}{\lambda_2+\theta})
\end{align}

Here, $\theta=P/2+\kappa_2$. This results in

\begin{align}
    |c_3'(t)|^2 \approx g_2^2 \frac{e^{-(P+\gamma+\kappa_1)t/2}}{g_1^2+\kappa_2^2}
\end{align}

Thus, $P_T=2\kappa_2\int_{0}^\infty |c_3'(t)|^2 dt$ becomes:

\begin{align}
    P_T=\frac{4 \kappa_2g_2^2}{(\gamma+\kappa_1)(\kappa_2^2+g_1^2)}=\frac{\eta}{100}
\end{align}

Note that we have ignored $P$ in the denominator because of its small magnitude compared to $\kappa_1,\gamma$. 
\nocite{*}

\bibliography{apssamp}

\end{document}